\documentclass[aps,prl,twocolumn,groupedaddress,amsmath,amssymb]{revtex4}
\usepackage{graphicx}% Include figure files
\usepackage{dcolumn}% Align table columns on decimal point
\usepackage{bm}% bold math

\begin{document}

%\preprint{APS/123-QED}

\title{Non-thermal origin of nonlinear transport across magnetically induced superconductor-metal-insulator transition}

\author{Y. Seo$^1$}
\altaffiliation[Present address:]{ Nano Science and Technology,
Sejong University, Seoul, Korea.}
\author{Y. Qin$^1$}
\author{C. L. Vicente$^1$}
\author{K. S. Choi$^{1,2}$}
\author{Jongsoo Yoon$^1$}
\affiliation{$^1$Department of Physics, University of Virginia,
Charlottesville, VA 22903, U.S.A. \\$^2$Department of Physics,
Sunchon National University, Sunchon, Jeonnam, Korea.}
 %\affiliation{$^2$Department of
%Physics, Sunchon National University, Sunchon, Jeonnam, Korea.}




\date{\today}% It is always \today, today,

\begin{abstract}
We have studied the effect of perpendicular magnetic fields and
temperatures on the nonlinear electronic transport in amorphous Ta
superconducting thin films. The films exhibit a magnetic field
induced metallic behavior intervening the superconductor-insulator
transition in the zero temperature limit. We show that the nonlinear
transport in the superconducting and metallic phase is of
non-thermal origin and accompanies an extraordinarily long voltage
response time.
\end{abstract}

\maketitle

In recent years, the suppression of superconductivity in
two-dimensions (2D) by means of increasing disorder (usually
controlled by film thickness) or applying magnetic fields has been a
focus of attention. Conventional treatments \cite{r1,r2,r3,r4,r5} of
electronic transport predict that in 2D the suppression of the
superconductivity leads to a direct superconductor-insulator
transition (SIT) in the limit of zero temperature (T = 0). This
traditional view has been challenged by the observation of magnetic
field (B) induced metallic behavior in amorphous MoGe
\cite{r6,r7,r8} and Ta thin films \cite{r9}. The unexpected metallic
behavior, intervening the $B$-driven SIT, is characterized by a drop
in resistance ($\rho$) followed by a saturation to a finite value as
$T \rightarrow 0$. The metallic resistance can be orders of
magnitude smaller than the normal state resistance ($\rho_n$)
implying that the metallic state exists as a separate phase rather
than a point in the phase diagram. Despite many theoretical
treatments \cite{r8,r10,r11,r12,r13,r14,r15,r16,r17}, a consensus on
the mechanism behind the metallic behavior is yet to be reached.
Proposed origins of the metallic behavior include bosonic
interactions in the non-superconducting phase\cite{r10,r11},
contribution of fermionic quasiparticles to the
conduction\cite{r12,r13}, and quantum phase
fluctuations\cite{r14,r15}.

In a recent paper \cite{r9} on the magnetically induced metallic
behavior in Ta films, we have reported the nonlinear voltage-current
($I$-$V$) characteristics that can be used to identify each phase.
The superconducting phase is unique in having both a hysteretic
$I$-$V$ and an ``immeasurably" small voltage response to currents
below an apparent critical current $I_c$. The metallic phase can be
identified by a differential resistance ($dV/dI$) that increases
with increasing $I$, whereas the insulating phase is identified by a
$dV/dI$ that decreases with increasing $I$. The contrasting
nonlinear $I$-$V$ in the metallic and insulating phase are shown in
Fig. 1(a).

%put figure 1 here

The main purpose of this Letter is to report that the origin of the
nonlinear transport, particularly in the superconducting and
metallic phase, is not a simple reflection of $T$-dependence of
$\rho$ via the unavoidable Joule heating. We describe the effect of
$B$ and $T$ on the nonlinear transport on which this conclusion is
based. We also present our studies on dynamic voltage response which
reveal strikingly long voltage response times that accompany the
nonlinear transport.

Our samples are dc sputter deposited Ta thin films on Si substrates.
The sputter chamber is baked at $\sim 110 ^\circ$C for several days,
reaching a base pressure of ~ 10-8 Torr. The chamber and Ta source
were cleaned by pre-sputtering for $\sim 30$ min at a rate of $\sim
1$nm/s. Films are grown at a rate of $\sim 0.05$ nm/s at an Ar
pressure of $\sim 4$ mTorr. Using a rotatable substrate holder up to
12 films, each with a different thickness, can be grown without
breaking the vacuum. In order to facilitate four point measurements,
the samples are patterned into a bridge (1 mm wide and 5 mm long)
using a shadow mask. Even though there were noticeable batch to
batch variations, the degree of disorder (evidenced by the values of
$\rho_n$) for films of the same batch increases monotonically with
decreasing film thickness. The superconducting properties of the
films are characteristic of homogeneously disordered thin films
\cite{r18}, and consistent with the results of x-ray structural
investigations \cite{r9}. The data presented in this paper are from
5 films grown in 4 batches. Parameters of the films are summarized
in Table \ref{tab:table1}.

\begin{table}
\caption{\label{tab:table1}List of sample parameters: nominal film
thickness, mean field $T_c$ at $B = 0$, normal state resistivity at
$4.2$ K, critical magnetic field at which the resistance reaches $90
\%$ of the high field saturation value, and correlation length
calculated from $\xi=\sqrt{\Phi_0/2\pi B_c}$  where $\Phi_0$ is the
flux quantum.}
\begin{ruledtabular}
\begin{tabular}{ccccccc}

 sample & batch &$t (nm)$ &$T_c (K)$&$\rho_n (\Omega / \square)$ &$B_c$ &$\xi$      \\
\hline

Ta 1& 1 & 3.5 & 0.584 & 1769
& 0.72 & 21  \\
Ta 2& 1 & 5.0 & 0.675 & 1180
& 0.88 & 19 \\
Ta 3& 2 & 5.7 & 0.770 & 1056
& 0.9 & 19 \\
Ta 4& 3 & 5.0 & 0.598 & 770
& - & - \\
Ta 5& 4 & 36 & 0.995 &69
& 2.0 & 13 \\

\end{tabular}
\end{ruledtabular}
%\footnotetext[1]{Here's the first, from Ref.~\onlinecite{feyn54}.}
\end{table}

\begin{figure}
\includegraphics{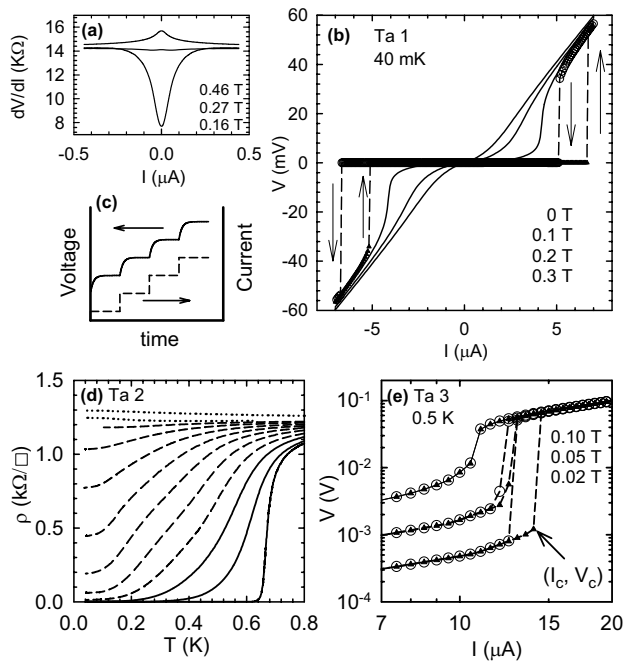}% Here is how to import EPS art
\caption{\label{fig:fig1} (a) $dV/dI$ vs. $I$ at 20 mK across the
metal-insulator boundary for a Ta film with $T_c$ = 0.23 K, adapted
from Ref.\cite{r9}. (b) Current biased I-V curves of sample Ta 1 at
40 mK and the indicated B. (c) Our method of $I$-$V$ measurements is
illustrated. (d) The $T$-dependence of $\rho$ for sample Ta 2 at $B$
= 0 - 1.0 T with 0.1 T interval and 3.0 T, measured at 7 Hz with a
current amplitude of 1 nA. The solid lines are to indicate the
superconducting phase in the low T limit, the dashed lines the
metallic phase, and the dotted lines insulating phase. (e) $I$-$V$
curves in $\log$-$\log$ scale for sample Ta 3. Filled triangles
(open circles) are for current increasing (decreasing) branch. The
data density is reduced to make individual symbols visible. The
arrow marks the critical current $I_c$ and voltage $V_c$. }
\end{figure}

The evolution of the $I$-$V$ curves across the superconductor-metal
boundary at $40$ mK is shown in Fig. \ref{fig:fig1}(b) for sample Ta
1. The hysteretic $I$-$V$, unique to the superconducting phase, is
indicated by the dashed lines. As $I$ is increased, the
superconductivity is abruptly quenched at a well-defined critical
current $I_c$. As $I$ is decreased from above $I_c$, the
superconductivity suddenly appears at a different current $I_c'$ $<$
$I_c$. The hysteresis becomes smaller with increasing $B$, and
vanishes near 0.1 T as the system is driven into the metallic phase
[solid lines in Fig. \ref{fig:fig1}(b)]. Typical $T$-dependence of
$\rho$ at various $B$ is shown in Fig. \ref{fig:fig1}(d) for another
sample Ta 2. In this sample, which is less disordered than Ta 1, the
superconducting phase extends up to $\sim$ 0.2 T [solid lines in
Fig. \ref{fig:fig1}(d)] and the metallic behavior is observed at
higher $B$ up to $\sim$ 0.9 T (dashed lines). Hereafter,
``superconducting regime" refers to the transport regime where I-V
is hysteretic and ``metallic regime" to the regime with nonlinear
(and reversible) $I$-$V$ with increasing $dV/dI$ with increasing
$I$.

All our $I$-$V$ curves are constructed by plotting the steady state
voltage at each bias current that is changed in small discreet steps
as illustrated in Fig. \ref{fig:fig1}(c). In order to ensure that
the steady state is reached at each step, the voltage is monitored
every 50 ms for up to 55 s while the current is kept constant. The
magnitude of the voltage jump at $I_c$ (or $I_c'$), which could be
as large as several orders of magnitude, was almost independent of
the current step size in the range 5 - 100 nA. Even with our
smallest steps of 5 nA, no steady state with a voltage within the
range covered by the jump was observed.

\begin{figure}
\includegraphics{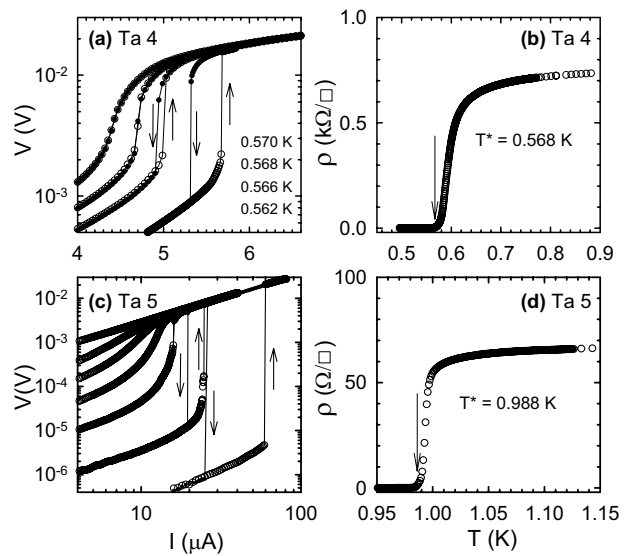}
\caption{\label{fig:fig2}(a) $I$-$V$ curves of sample Ta 4 at B = 0
and at the indicated $T$. Open (filled) circles are for the current
increasing (decreasing) branch. (b) $T$-dependence of $\rho$ for
sample Ta 4 at $B$ = 0. The arrow is to indicate the temperature T*
where the hysteresis vanishes. (c) The $I$-$V$ of the sample Ta 5 at
B = 0. The temperature of each trace is, from the top, 1.000, 0.994,
0.992, 0.990, 0.988, 0.984, and 0.976 K. (d) The $T$-dependence of
$\rho$ for sample Ta 5 at $B$ = 0. }
\end{figure}

Our investigations on how $B$ and $T$ influence the nonlinear
transport indicate that these quantities play similar roles. The
main effect of increasing $T$ is to lower the superconductor-metal
``critical" field $B_c ^{sm}$; the $B$-driven evolution of the
$I$-$V$ curves at an elevated $T$ [Fig. \ref{fig:fig1}(e)] remains
qualitatively the same as that in the low $T$ limit [Fig.
\ref{fig:fig1}(b)]. More importantly, the evolution of the $I$-$V$
curves as a function of $T$ [shown in Fig. \ref{fig:fig2}(a) and
(c)] is strikingly similar to that caused by $B$ [shown in Fig.
\ref{fig:fig1}(b) and (e)].

The field $B_c ^{sm}$ decreases with increasing $T$ and reaches zero
at a well-defined temperature $T^*$, which is close to $T_c$ as
shown by the arrows in Fig. \ref{fig:fig2}(b) and (d). This,
together with the observations described above, means that $B_c
^{sm}$ is a well-defined line in $B$-$T$ plane separating the
superconducting and metallic regime. We point out that, in terms of
nonlinear transport, the electronic properties at $B \gtrsim B_c
^{sm}$ in the low $T$ limit, where the unexpected metallic behavior
intervening SIT is observed, are indistinguishable from those at
high temperatures, for example at $T \gtrsim T^*$ and $B \approx 0$.

\begin{figure}
\includegraphics{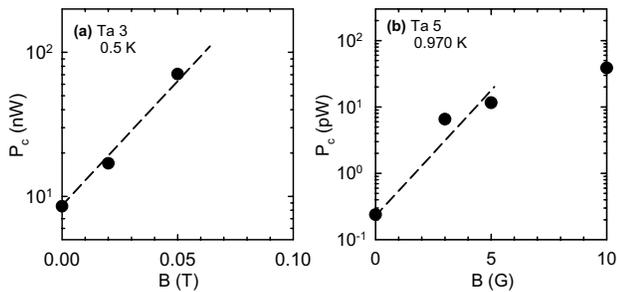}
\caption{\label{fig:fig3} (a) The $B$-dependence of the critical
power, $P_c$=$I_c$$V_c$, for sample Ta 3 at 0.5 K. $I_c$ and $V_c$
are marked by an arrow in Fig. 1(e). The dashed line is to guide an
eye. (b) $P_c$ vs. $B$ plot for sample Ta 5 at 0.970 K.}
\end{figure}

An important finding from our investigations is that the voltage
jump in the superconducting regime is of non-thermal origin. If
Joule heating is significant, the electron temperature $T_e$ would
be determined by the balance of the Joule heating power and the heat
drain rate to the stage where the sample is thermally anchored. A
large heating power at a high bias current could make $T_e$
substantially higher than the stage temperature. If $T_e$ reaches
near $T_c$ where $\rho$ sharply rises with $T$, an increase in $T_e$
could cause an increase in the heating power, which in turn causes a
further increase in $T_e$. Such a positive self-feedback would make
$T_e$ unstable and run away beyond $T_c$, resulting in a sudden
quenching of the superconductivity appearing as a voltage jump. This
scenario can be tested by applying weak $B$. The magnetic fields
lower $T_c$ while the net thermal conductance between the sample and
the stage would remain almost unaffected. Therefore, in the heating
scenario the critical power, $P_c=I_cV_c$ where $I_c$ and $V_c$ are
the current and voltage at the onset of the voltage jump on the
current increasing branch [marked by an arrow in Fig.
\ref{fig:fig1}(e)], is expected to be weakly decreasing function of
$B$. However, as shown in Fig. \ref{fig:fig3}(a) and (b), $P_c$ is
found to increase by an order of magnitude or more under weak $B$.
This clearly demonstrates that the voltage jump in the
superconducting regime has a non-thermal origin. We note that $V_c$
is the highest steady state voltage in the superconducting state
before the current-induced sudden quenching of the
superconductivity. In repeated runs after thermal cyclings to above
10 K, the value of $V_c$ was reproducible within several percent
even with different current step sizes in the range 5 - 100 nA.

The $I$-$V$ curves in Fig. \ref{fig:fig1}-\ref{fig:fig2} clearly
show that the discontinuity in $I$-$V$ in the superconducting regime
evolves into the point of the largest slope in the continuous
$I$-$V$ in the metallic regime. This strongly suggests that the
sudden quenching of the superconductivity at $I_c$ and the nonlinear
transport in the metallic regime are caused by the same mechanism,
which has been shown above to be of non-thermal origin.

\begin{figure}
\includegraphics{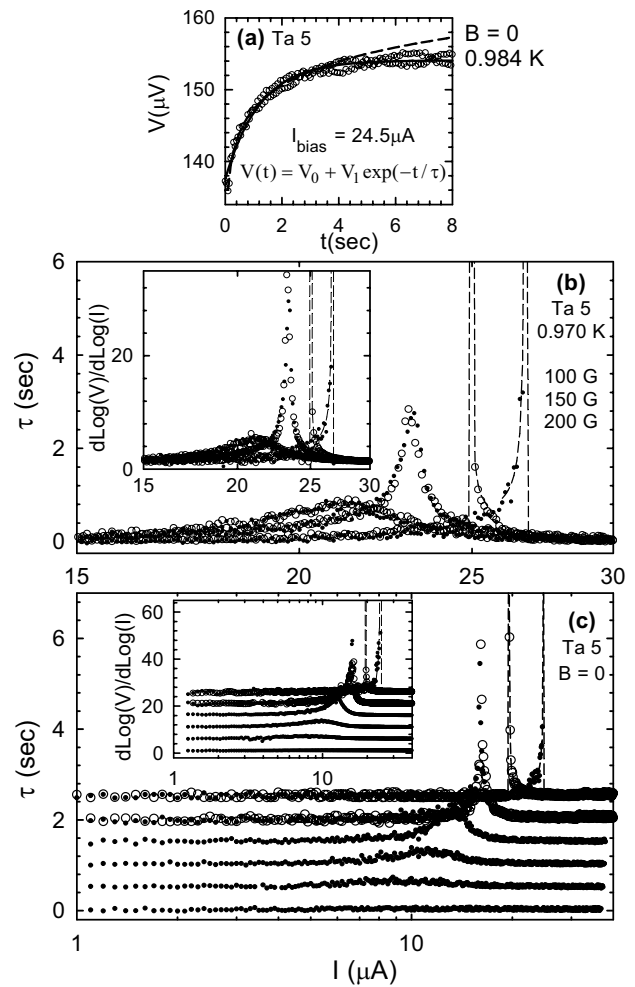}
\caption{\label{fig:fig4}(a) An example of $V$-$t$ trace at $I$ =
24.5 A. The solid line is a fit to the exponential function shown.
The dashed line is a fit to a logarithmic $t$-dependence for the
first 3 s. (b) $t$ vs. $I$ plot at the indicated $B$ and $T$. Filled
(open) circles are for current increasing (decreasing) branch. The
dashed lines are to guide an eye. Inset: the slope of $I$-$V$ curves
in $\log$-$\log$ scale at the same $B$ and $T$. (c) $\tau$-$I$ plot
at $B$ = 0. The temperature of each trace is, from the top, 0.984,
0.988, 0.990, 0.992, 0.994, and 1.000 K. Each trace is vertically
shifted successively. For the bottom four traces, only current
increasing branch is shown. Inset: $d(\log V)$/$d(\log I)$ is
plotted at the same $T$ and $B$. Each trace is vertically shifted
successively. }
\end{figure}

Now we turn to the discussion on the dynamic voltage response. The
dynamic voltage response was studied by analyzing the voltage-time
($V$-$t$) traces to determine how fast the steady state is reached
at each bias current. As shown by the solid line in Fig.
\ref{fig:fig4}(a), the $t$-dependence is well described by an
exponential function. The parameters $V_0$ and $V_1$ are determined
from the measured steady state voltage and $V(t=0)$ which is the
steady state voltage at the previous bias current. The parameter
$\tau$ is defined as the voltage response time constant, and
obtained from a least squared error fitting procedure.

Figure \ref{fig:fig4}(b) shows the $\tau$-$I$ plots at three
different $B$ across the superconductor-metal boundary. In the
superconducting regime ($B$=100 G), the time constant exhibits a
hysteretic and diverging behavior with approaching $I_c$ from below
($I_c'$ from above) where the $I$-$V$ is discontinuous. In the
metallic regime (150 and 200 G) where $I$-$V$ is continuous and
reversible, the $\tau$-traces are also continuous and reversible.
However, a prominent peak structure is evident in the data.
Interestingly, the peak or diverging feature in $\tau$ almost
exactly coincides with the nonlinear transport. The inset shows the
slopes of $I$-$V$ curves in $\log$-$\log$ scale. At current where
$\tau$ is large the transport is nonlinear $[d(\log V)/d(\log I) >
1]$, and at current where $\tau$ is almost zero the transport is
linear $[d(\log V)/d(\log I) = 1]$. Qualitatively the same $\tau$
behavior is observed when the superconductor-metal boundary is
crossed by increasing $T$. This is shown in Fig. \ref{fig:fig4}(c).
Note that the traces in Fig. \ref{fig:fig4}(c) and the inset are
successively shifted vertically.

It is surprising to find that the time constant in the peak or
diverging region is as long as several seconds. We emphasize that
the long time constant is not due to instrumentation. This is best
demonstrated by the systematics of the data. Changing $B$ or $T$
systematically shifts the peak or divergence while outside the
narrow peak or diverging region the time constants remain almost
zero. The time constants were measured to be the same within the
scatter of the data for current steps of 5 nA (not shown) and 100 nA
(shown).

In the past, hysteresis accompanying a long response time has been
studied in the context of irreversible magnetic properties in type
II superconductors \cite{r19}. Large magnetic relaxation rates
observed in magnetization measurements \cite{r20} are believed to
arise from thermal activation of magnetic flux lines out of pinning
sites. The depinning process leads to a redistribution of flux lines
causing a change in magnetization with time. Although our hysteresis
with a long response time is observed in transport, not
magnetization measurements, it still may be possible to understand
in terms of pinning-depinning of vortices. A pinning-depinning
transition arises from the competition between disorder-induced
pinning force and Lorentz driving force due to the bias current.
Under such a competition, the vortex motion is analogous to the flow
of sand grains in a sand pile \cite{r21}, where the competition is
between the jamming due to the granularity of the system and
gravitational force. Indeed, hysteresis \cite{r22} and slow
relaxation rates \cite{r23} have been observed in granular flow
under mechanical vibrations. Logarithmic time dependence has been
observed in measurements of relaxation of magnetization \cite{r19}
and experiments on granular flow\cite{r23}. The dashed line in Fig.
\ref{fig:fig4}(a) is the fit of our data to a logarithmic time
dependence, and describes the data well for t $<$ 3 s. Nevertheless,
over the entire range of the data, the exponential function (solid
line) fits our data better.

Finally, we briefly discuss an interesting implication of the
unusually long response time accompanying the nonlinear transport at
$B=0$ and $T \gtrsim T^*$. Nonlinear transport of 2D superconductors
at B = 0 is usually understood in the framework of
Kosterlitz-Thouless (KT) theory \cite{r24}, where the
superconducting transition corresponds to a thermodynamic
instability of vortex-antivortex pairs in 2D. In this picture,
current-induced dissociation of vortex pairs in the superconducting
phase is expected to lead to nonlinear transport in the fashion, $V
\propto I^\textrm{a}$ with a $>$ 3 \cite{r25}. It has been argued
\cite{r26} that in a real system finite size effect can induce free
vortices altering the power law $I$-$V$. The resulting $I$-$V$
curves obtained in numerical simulations \cite{r26} show a
pronounced peak structure in $d(\log V)/d(\log I)$ resembling those
shown in the inset of Fig. \ref{fig:fig4}(c). However, the voltage
response time which is observed to be as long as several seconds, is
too long to be reasonable with the KT framework where the vortex
dissociation is expected to occur in the time scale of quasiparticle
scattering, typically $\sim 10^{-9}$s \cite{r27}. At present,
whether dynamics of KT vortices in the presence of disorder can have
such long time scales is not clear, and the understanding of the
nonlinear transport requires the development of a theoretical
framework for nonequilibrium dynamics.

In summary, we have shown the magnetic field and temperature driven
evolution of phase-identifying nonlinear $I$-$V$ characteristics of
Ta thin films that exhibit an unexpected metallic phase intervening
SIT in the low $T$ limit. Our observations indicate that a
non-thermal mechanism is behind both the nonlinear transport in the
metallic phase and sudden current-induced quenching of the
superconductivity in the superconducting phase. Our dynamic voltage
response studies suggest a possible link of the metallic behavior to
the dynamics of vortices in the presence of disorder.

\begin{acknowledgments}
Authors acknowledge fruitful discussions with V. Galitski, H.
Fertig, and E. Kolomeisky. This work is supported by NSF.
\end{acknowledgments}

\bibliography{manu}% Produces the bibliography via BibTeX.

\begin{thebibliography}{27}
\expandafter\ifx\csname natexlab\endcsname\relax\def\natexlab#1{#1}\fi
\expandafter\ifx\csname bibnamefont\endcsname\relax
  \def\bibnamefont#1{#1}\fi
\expandafter\ifx\csname bibfnamefont\endcsname\relax
  \def\bibfnamefont#1{#1}\fi
\expandafter\ifx\csname citenamefont\endcsname\relax
  \def\citenamefont#1{#1}\fi
\expandafter\ifx\csname url\endcsname\relax
  \def\url#1{\texttt{#1}}\fi
\expandafter\ifx\csname urlprefix\endcsname\relax\def\urlprefix{URL }\fi
\providecommand{\bibinfo}[2]{#2}
\providecommand{\eprint}[2][]{\url{#2}}

\bibitem[{\citenamefont{Abrahams et~al.}(1979)\citenamefont{Abrahams, Anderson,
  Licciardello, and Ramakrishnan}}]{r1}
\bibinfo{author}{\bibfnamefont{E.}~\bibnamefont{Abrahams}},
  \bibinfo{author}{\bibfnamefont{P.~W.} \bibnamefont{Anderson}},
  \bibinfo{author}{\bibfnamefont{D.~C.} \bibnamefont{Licciardello}},
  \bibnamefont{and} \bibinfo{author}{\bibfnamefont{T.~V.}
  \bibnamefont{Ramakrishnan}}, \bibinfo{journal}{Phys.\ Rev.\ Lett.}
  \textbf{\bibinfo{volume}{42}}, \bibinfo{pages}{673} (\bibinfo{year}{1979}).

\bibitem[{\citenamefont{Finkelshtein}(1987)}]{r2}
\bibinfo{author}{\bibfnamefont{A.}~\bibnamefont{Finkelshtein}},
  \bibinfo{journal}{JETP\ Lett.} \textbf{\bibinfo{volume}{45}},
  \bibinfo{pages}{46} (\bibinfo{year}{1987}).

\bibitem[{\citenamefont{Larkin}(1999)}]{r3}
\bibinfo{author}{\bibfnamefont{A.}~\bibnamefont{Larkin}},
  \bibinfo{journal}{Ann.\ Phys.\ (Leipzig)} \textbf{\bibinfo{volume}{8}},
  \bibinfo{pages}{785} (\bibinfo{year}{1999}).

\bibitem[{\citenamefont{Fisher}(1990)}]{r4}
\bibinfo{author}{\bibfnamefont{M.~P.~A.} \bibnamefont{Fisher}},
  \bibinfo{journal}{Phys.\ Rev.\ Lett.} \textbf{\bibinfo{volume}{65}},
  \bibinfo{pages}{923} (\bibinfo{year}{1990}).

\bibitem[{\citenamefont{Fisher et~al.}(1990)\citenamefont{Fisher, Grinstein,
  and Girvin}}]{r5}
\bibinfo{author}{\bibfnamefont{M.~P.~A.} \bibnamefont{Fisher}},
  \bibinfo{author}{\bibfnamefont{G.}~\bibnamefont{Grinstein}},
  \bibnamefont{and} \bibinfo{author}{\bibfnamefont{S.~M.}
  \bibnamefont{Girvin}}, \bibinfo{journal}{Phys.\ Rev.\ Lett.}
  \textbf{\bibinfo{volume}{64}}, \bibinfo{pages}{587} (\bibinfo{year}{1990}).

\bibitem[{\citenamefont{Mason and Kapitulnik}(1999)}]{r6}
\bibinfo{author}{\bibfnamefont{N.}~\bibnamefont{Mason}} \bibnamefont{and}
  \bibinfo{author}{\bibfnamefont{A.}~\bibnamefont{Kapitulnik}},
  \bibinfo{journal}{Phys.\ Rev.\ Lett.} \textbf{\bibinfo{volume}{82}},
  \bibinfo{pages}{5371} (\bibinfo{year}{1999}).

\bibitem[{\citenamefont{Mason and Kapitulnik}(2002)}]{r7}
\bibinfo{author}{\bibfnamefont{N.}~\bibnamefont{Mason}} \bibnamefont{and}
  \bibinfo{author}{\bibfnamefont{A.}~\bibnamefont{Kapitulnik}},
  \bibinfo{journal}{Phys.\ Rev.\ B} \textbf{\bibinfo{volume}{65}},
  \bibinfo{pages}{220505(R)} (\bibinfo{year}{2002}).

\bibitem[{\citenamefont{Ephron et~al.}(1996)\citenamefont{Ephron, Yazdani,
  Kapitulnik, and Beasley}}]{r8}
\bibinfo{author}{\bibfnamefont{D.}~\bibnamefont{Ephron}},
  \bibinfo{author}{\bibfnamefont{A.}~\bibnamefont{Yazdani}},
  \bibinfo{author}{\bibfnamefont{A.}~\bibnamefont{Kapitulnik}},
  \bibnamefont{and} \bibinfo{author}{\bibfnamefont{M.~R.}
  \bibnamefont{Beasley}}, \bibinfo{journal}{Phys.\ Rev.\ Lett}
  \textbf{\bibinfo{volume}{76}}, \bibinfo{pages}{1529} (\bibinfo{year}{1996}).

\bibitem[{\citenamefont{Qin et~al.}(2006)\citenamefont{Qin, Vicente, and
  Yoon}}]{r9}
\bibinfo{author}{\bibfnamefont{Y.}~\bibnamefont{Qin}},
  \bibinfo{author}{\bibfnamefont{C.~L.} \bibnamefont{Vicente}},
  \bibnamefont{and} \bibinfo{author}{\bibfnamefont{J.}~\bibnamefont{Yoon}},
  \bibinfo{journal}{Phys.\ Rev.\ B} \textbf{\bibinfo{volume}{73}},
  \bibinfo{pages}{100505(R)} (\bibinfo{year}{2006}).

\bibitem[{\citenamefont{Dalidovich and Phillips}(2001)}]{r10}
\bibinfo{author}{\bibfnamefont{D.}~\bibnamefont{Dalidovich}} \bibnamefont{and}
  \bibinfo{author}{\bibfnamefont{P.}~\bibnamefont{Phillips}},
  \bibinfo{journal}{Phys.\ Rev.\ B} \textbf{\bibinfo{volume}{64}},
  \bibinfo{pages}{052507} (\bibinfo{year}{2001}).

\bibitem[{\citenamefont{Dalidovich and Phillips}(2002)}]{r11}
\bibinfo{author}{\bibfnamefont{D.}~\bibnamefont{Dalidovich}} \bibnamefont{and}
  \bibinfo{author}{\bibfnamefont{P.}~\bibnamefont{Phillips}},
  \bibinfo{journal}{Phys.\ Rev.\ Lett} \textbf{\bibinfo{volume}{89}},
  \bibinfo{pages}{027001} (\bibinfo{year}{2002}).

\bibitem[{\citenamefont{Galitski et~al.}(2005)\citenamefont{Galitski, Refael,
  Fisher, and Senthil}}]{r12}
\bibinfo{author}{\bibfnamefont{V.~M.} \bibnamefont{Galitski}},
  \bibinfo{author}{\bibfnamefont{G.}~\bibnamefont{Refael}},
  \bibinfo{author}{\bibfnamefont{M.~P.~A.} \bibnamefont{Fisher}},
  \bibnamefont{and} \bibinfo{author}{\bibfnamefont{T.}~\bibnamefont{Senthil}},
  \bibinfo{journal}{Phys.\ Rev.\ Lett} \textbf{\bibinfo{volume}{95}},
  \bibinfo{pages}{077002} (\bibinfo{year}{2005}).

\bibitem[{\citenamefont{Kapitulnik et~al.}(2001)\citenamefont{Kapitulnik,
  Mason, Kivelson, and Chakravarty}}]{r13}
\bibinfo{author}{\bibfnamefont{A.}~\bibnamefont{Kapitulnik}},
  \bibinfo{author}{\bibfnamefont{N.}~\bibnamefont{Mason}},
  \bibinfo{author}{\bibfnamefont{S.~A.} \bibnamefont{Kivelson}},
  \bibnamefont{and}
  \bibinfo{author}{\bibfnamefont{S.}~\bibnamefont{Chakravarty}},
  \bibinfo{journal}{Phys.\ Rev.\ B} \textbf{\bibinfo{volume}{63}},
  \bibinfo{pages}{125322} (\bibinfo{year}{2001}).

\bibitem[{\citenamefont{Das and Doniach}(2001)}]{r14}
\bibinfo{author}{\bibfnamefont{D.}~\bibnamefont{Das}} \bibnamefont{and}
  \bibinfo{author}{\bibfnamefont{S.}~\bibnamefont{Doniach}},
  \bibinfo{journal}{Phys.\ Rev.\ B} \textbf{\bibinfo{volume}{64}},
  \bibinfo{pages}{134511} (\bibinfo{year}{2001}).

\bibitem[{\citenamefont{Spivak et~al.}(2001)\citenamefont{Spivak, Zyuzin, and
  Hruska}}]{r15}
\bibinfo{author}{\bibfnamefont{B.}~\bibnamefont{Spivak}},
  \bibinfo{author}{\bibfnamefont{A.}~\bibnamefont{Zyuzin}}, \bibnamefont{and}
  \bibinfo{author}{\bibfnamefont{M.}~\bibnamefont{Hruska}},
  \bibinfo{journal}{Phys.\ Rev.\ B} \textbf{\bibinfo{volume}{64}},
  \bibinfo{pages}{132502} (\bibinfo{year}{2001}).

\bibitem[{\citenamefont{Tewari}(2004)}]{r16}
\bibinfo{author}{\bibfnamefont{S.}~\bibnamefont{Tewari}},
  \bibinfo{journal}{Phys.\ Rev.\ B} \textbf{\bibinfo{volume}{69}},
  \bibinfo{pages}{014512} (\bibinfo{year}{2004}).

\bibitem[{\citenamefont{Ng and Lee}(2001)}]{r17}
\bibinfo{author}{\bibfnamefont{T.~K.} \bibnamefont{Ng}} \bibnamefont{and}
  \bibinfo{author}{\bibfnamefont{D.~K.} \bibnamefont{Lee}},
  \bibinfo{journal}{Phys.\ Rev.\ B} \textbf{\bibinfo{volume}{64}},
  \bibinfo{pages}{144509} (\bibinfo{year}{2001}).

\bibitem[{\citenamefont{Goldman and Markovic}(1998)}]{r18}
\bibinfo{author}{\bibfnamefont{A.~M.} \bibnamefont{Goldman}} \bibnamefont{and}
  \bibinfo{author}{\bibfnamefont{N.}~\bibnamefont{Markovic}},
  \bibinfo{journal}{Phys.\ Today} \textbf{\bibinfo{volume}{49}},
  \bibinfo{pages}{39} (\bibinfo{year}{1998}).

\bibitem[{\citenamefont{Yeshurun et~al.}(1996)\citenamefont{Yeshurun,
  Malozemoff, and Shaulov}}]{r19}
\bibinfo{author}{\bibfnamefont{Y.}~\bibnamefont{Yeshurun}},
  \bibinfo{author}{\bibfnamefont{A.~P.} \bibnamefont{Malozemoff}},
  \bibnamefont{and} \bibinfo{author}{\bibfnamefont{A.}~\bibnamefont{Shaulov}},
  \bibinfo{journal}{Rev.\ Mod.\ Phys.} \textbf{\bibinfo{volume}{68}},
  \bibinfo{pages}{911} (\bibinfo{year}{1996}).

\bibitem[{\citenamefont{Beasley et~al.}(1969)\citenamefont{Beasley, Labusch,
  and Webb}}]{r20}
\bibinfo{author}{\bibfnamefont{M.~R.} \bibnamefont{Beasley}},
  \bibinfo{author}{\bibfnamefont{R.}~\bibnamefont{Labusch}}, \bibnamefont{and}
  \bibinfo{author}{\bibfnamefont{W.~W.} \bibnamefont{Webb}},
  \bibinfo{journal}{Phys.\ Rev.} \textbf{\bibinfo{volume}{181}},
  \bibinfo{pages}{682} (\bibinfo{year}{1969}).

\bibitem[{\citenamefont{Pla and Nori}(1991)}]{r21}
\bibinfo{author}{\bibfnamefont{S.}~\bibnamefont{Pla}} \bibnamefont{and}
  \bibinfo{author}{\bibfnamefont{F.}~\bibnamefont{Nori}},
  \bibinfo{journal}{Phys.\ Rev.\ Lett} \textbf{\bibinfo{volume}{67}},
  \bibinfo{pages}{919} (\bibinfo{year}{1991}).

\bibitem[{\citenamefont{Tennakoon et~al.}(1999)\citenamefont{Tennakoon, Kondic,
  and Behringer}}]{r22}
\bibinfo{author}{\bibfnamefont{S.~G.~K.} \bibnamefont{Tennakoon}},
  \bibinfo{author}{\bibfnamefont{L.}~\bibnamefont{Kondic}}, \bibnamefont{and}
  \bibinfo{author}{\bibfnamefont{R.~P.} \bibnamefont{Behringer}},
  \bibinfo{journal}{Europhys.\ Lett.} \textbf{\bibinfo{volume}{45}},
  \bibinfo{pages}{470} (\bibinfo{year}{1999}).

\bibitem[{\citenamefont{Jaeger et~al.}(1988)\citenamefont{Jaeger, Liu, and
  Nagel}}]{r23}
\bibinfo{author}{\bibfnamefont{H.~M.} \bibnamefont{Jaeger}},
  \bibinfo{author}{\bibfnamefont{C.-H.} \bibnamefont{Liu}}, \bibnamefont{and}
  \bibinfo{author}{\bibfnamefont{S.}~\bibnamefont{Nagel}},
  \bibinfo{journal}{Europhys.\ Lett.} \textbf{\bibinfo{volume}{62}},
  \bibinfo{pages}{40} (\bibinfo{year}{1988}).

\bibitem[{\citenamefont{Kosterlitz and Thouless}(1973)}]{r24}
\bibinfo{author}{\bibfnamefont{J.~D.} \bibnamefont{Kosterlitz}}
  \bibnamefont{and} \bibinfo{author}{\bibfnamefont{D.~J.}
  \bibnamefont{Thouless}}, \bibinfo{journal}{J.\ Phys.\ C}
  \textbf{\bibinfo{volume}{6}}, \bibinfo{pages}{1181} (\bibinfo{year}{1973}).

\bibitem[{\citenamefont{Halperin and Nelson}(1979)}]{r25}
\bibinfo{author}{\bibfnamefont{B.~I.} \bibnamefont{Halperin}} \bibnamefont{and}
  \bibinfo{author}{\bibfnamefont{D.~R.} \bibnamefont{Nelson}},
  \bibinfo{journal}{J.\ Low Temp.\ Phys} \textbf{\bibinfo{volume}{36}},
  \bibinfo{pages}{599} (\bibinfo{year}{1979}).

\bibitem[{\citenamefont{Medvedyeva et~al.}(2000)\citenamefont{Medvedyeva, Kim,
  and Minnhagen}}]{r26}
\bibinfo{author}{\bibfnamefont{K.}~\bibnamefont{Medvedyeva}},
  \bibinfo{author}{\bibfnamefont{B.}~\bibnamefont{Kim}}, \bibnamefont{and}
  \bibinfo{author}{\bibfnamefont{P.}~\bibnamefont{Minnhagen}},
  \bibinfo{journal}{Phys.\ Rev.\ B} \textbf{\bibinfo{volume}{62}},
  \bibinfo{pages}{14531} (\bibinfo{year}{2000}).

\bibitem[{\citenamefont{Ruck et~al.}(2000)\citenamefont{Ruck, Trodahl, Abele,
  and Geselbracht}}]{r27}
\bibinfo{author}{\bibfnamefont{B.~J.} \bibnamefont{Ruck}},
  \bibinfo{author}{\bibfnamefont{H.~J.} \bibnamefont{Trodahl}},
  \bibinfo{author}{\bibfnamefont{J.~C.} \bibnamefont{Abele}}, \bibnamefont{and}
  \bibinfo{author}{\bibfnamefont{M.~J.} \bibnamefont{Geselbracht}},
  \bibinfo{journal}{Phys.\ Rev.\ B} \textbf{\bibinfo{volume}{62}},
  \bibinfo{pages}{12468} (\bibinfo{year}{2000}).

\end{thebibliography}

\end{document}